\documentclass[pra,superscriptaddress,showpacs,twocolumn,a4paper]{revtex4}
\usepackage{latexsym,graphicx,amsmath,amsfonts,enumerate}

\def\op#1{\hat{#1}}
\def\ket#1{| #1 \rangle}

\def\bra#1{\langle #1 |}

\def\proj#1{\ket{#1}\bra{#1}}

\def\inprod#1#2{\langle #1 |  #2 \rangle}
\def\op#1#2{|#1\rangle\langle#2|}

\def\one{\mathbb{I}}
\def\tr{\text{Tr}}

\begin{document}

\title{Binary search trees for generalized measurement}

\author{Erika Andersson} \affiliation{SUPA, Department of Physics, University
  of Strathclyde, Glasgow G4 0NG, United Kingdom} \affiliation{SUPA, School of
  Engineering and Physical Sciences, Heriot-Watt University, Edinburgh EH14
  4AS, United Kingdom} \author{Daniel~K.~L.~\surname{Oi}}
%\email{Daniel.Oi@strath.ac.uk}
\affiliation{SUPA, Department of Physics, University of Strathclyde, Glasgow
  G4 0NG, United Kingdom} \date{17th December 2007}

\begin{abstract}
  Generalized quantum measurements (POVMs or POMs) are important for optimally
  extracting information for quantum communication and computation. The
  standard realization via the Neumark extension requires extensive resources
  in the form of operations in an extended Hilbert space. For an arbitrary
  measurement, we show how to construct a binary search tree with a depth
  logarithmic in the number of possible outcomes. This could be implemented
  experimentally by coupling the measured quantum system to a probe qubit
  which is measured, and then iterating.
\end{abstract}


\maketitle

\section{Introduction}

A crucial element of quantum information processing (QIP) and communication is
measurement of a quantum system to access its information content, hence
determining optimal measurements is important. As QIP steps out from the
pages of theory and into the laboratory, one has to find
implementations given the usual limited resources of the real world. The most
general measurement one can perform on a quantum system is given by a positive
operator valued measure (POVM)~\cite{helstrom} which can be considered as a
projective measurement on an extended Hilbert space of which the original
state resides on a (proper) sub-space. Its realization via the Neumark
extension~\cite{neumark, peres} requires, broadly speaking, that the extended
Hilbert space should have as many dimensions as there are possible outcomes of the
POVM. This has been described for atoms or ions~\cite{atompom1, atompom2} and
for linear optics~\cite{hillerypom}, and POVMs have been realised on optically
encoded quantum information~\cite{huttner, clarke1, clarke2, steinberg}.

For many physical systems, however, it is difficult or impossible to find
enough extra dimensions, let alone perform operations on the extended
system, hence a more efficient method is desirable. Recently, a method was
discovered which allows an arbitrary POVM to be performed by adding only a
single extra dimension to a system, essentially checking the measurement
outcomes one by one~\cite{WY2006}. However, when the number of possible
measurement outcomes becomes large, more time efficient measurement
strategies, also carrying a minimal dimensional overhead, would be useful.  It
is clear that a sequence of partial conditional measurements implements a
final effective POVM with many elements~\cite{peres, GP2006}. Here we show,
given any POVM, how to construct a suitable binary search
tree of two-outcome POVMs by coupling the original system with a single
qubit~\footnote{ This ``obvious'' result has been claimed
  before~\cite{QNWE1999, OT2005} though we have been unable to find a
  constructive proof of its universality in the existing literature.
  In~\cite{LV2001}, a partial result similar to Eq.(\ref{eq:soln}) was stated,
  but a linear search protocol implied.}. This way, a measurement with $N=2^t$
outcomes can be implemented in $t$ steps, resulting in a significant speedup.
Existing experimental realizations could easily be adapted to this
method~\cite{guerlin2007}.

\section{Generalized Measurement}

A quantum measurement is often considered to be a projection in a complete
basis of the $d$-dimensional Hilbert space. However, many experimental
measurements are not well described by this. More generally, we only require
of a measurement that the outcome probabilities are positive and sum to one,
and satisfy convex linearity over mixtures of states. This leads to the
framework of generalized quantum measurements, where a measurement is
represented by a set of positive operators $\{M_j\}_{j=1}^N,\text{
}\bra{\psi}M_j\ket{\psi}\ge 0\text{ }\forall \ket{\psi}$, which sum to unity,
$\sum_j M_j=\one$. Each outcome $j$ is associated with an operator $M_j$, and
occurs with probability $ %\text{Pr}(j)
p_j=\tr[M_j\rho]$, where $\rho$ is the measured state. Hence a generalized
measurement is usually called a positive operator valued measure (POVM) or
probability operator measure (POM).

The Neumark extension provides a way of performing any POVM via
projective measurements, albeit in an extended space. Without loss of
generality, assume that each measurement operator $M_j$ is proportional to a
one-dimensional projector $M_j=\proj{\psi_j}$ where $\ket{\psi_j}$ is not
neccessarily normalized~\footnote{Otherwise we can expand it as a sum of such
  terms and group the outcomes together.}.  If $N$ is the number of outcomes
and $d$ the dimension of the Hilbert space, then $N\ge d$ will hold. If we
form the $d\times N$ rectangular array $(M)_{jk}=\inprod{k}{\psi_j}$, where
$\{\ket{k}\}$ is the computational basis, then the completeness relation
implies that the columns of $(M)$ are orthonormal $N$-dimensional
vectors. Hence $(M)$ can be completed to an $N$-dimensional unitary matrix
$U_M$ whose $j^{th}$ row represents a state $|\psi_j^{ext}\rangle$ in an
$N$-dimensional extended Hilbert space.  The normalized projector
$|\psi_j^{ext}\rangle\langle\psi_j^{ext}|$ corresponds to outcome $j$ for the
original system. This procedure corresponds to applying $U_M^\dagger$
to the extended Hilbert space in which the original system is embedded, and
then making a projective measurement in the computational basis. If $N$ is
large, it may be infeasible to manipulate the required extended quantum system
all at once, and we will therefore look at a way to reduce the number of
ancillary dimensions by making sequential measurements.

\section{Sequential Measurement}

The $\{M_j\}$ are sufficient to determine the measurement probabilities, but
the post-measurement state is not uniquely defined.  However, for any
\emph{realization} we can find Kraus operators $\{m_j\}$, where
$M_j=m_j^\dagger m_j\text{ }\forall j$, which tell us how the quantum state is
affected~\cite{kraus,peres}. If outcome $j$ is obtained, then the quantum
state $\rho$ is transformed to $\rho_j = m_j \rho m_j^\dagger/\tr[m_j \rho
m_j^\dagger]$. A subsequent measurement, in general depending on the outcome
$j$, acts on this transformed state.

A series of measurements can be viewed as an effective single generalized
measurement, the sequence of outcomes determining the cumulative measurement
operator. If the sequence $j_1, j_2, ... ,j_t$ of outcomes have
$\{M_{j_1}^1,M_{j_2}^2,\ldots,M_{j_t}^t\}$ and $\{m_{j_k}^k\}$ as
corresponding measurement and Kraus operators, then the final
effective measurement operator and Kraus operator are given by
\begin{subequations}
\begin{eqnarray}
M_{j_1,j_2,\ldots,j_t}&=&m_{j_1,j_2,\ldots,j_t}^\dagger m_{j_1,j_2,\ldots,j_t},\\
m_{j_1,j_2,\ldots,j_t}&=&m_{j_t}^t m_{j_{t-1}}^{t-1}\ldots m_{j_1}^1.
\label{eq:seq}
\end{eqnarray}
\end{subequations}
Here we assume that the measurement operators and the Kraus operators are $d
\times d$ operators, i.e. the measurements map the system to a Hilbert space
of the same dimension~\footnote{This is not the case in general, in photon counting, 
for instance, photons are mapped to the vacuum state.}.  Hence,
a sequence of measurements requires non-destructive measurement, e.g. indirect
measurement of a system by measuring a probe after it has interacted with the
system.

\begin{figure}
\includegraphics[width=0.45\textwidth]{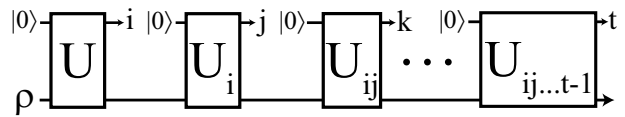}
\caption{Sequential POVM. A probe system in the state $|0\rangle$
  interacts with the measured system $\rho$ via unitaries. The probe is
  measured in the computational basis. Subsequent measurements, determined by
  $U_x$, are conditional upon the preceding results $i,j,k,\ldots$. The probe
  is reset before each measurement.}
\label{fig:seq}
\end{figure}

A series of binary outcome measurements is shown in Figure~\ref{fig:seq}. The
simplest probe is a two-level system (qubit), giving a binary measurement, a
$d$-level probe allowing a $d$-outcome measurement. A unitary operator couples
the probe with the system, e.g. via a coupling Hamiltonian over a set
period. This in general entangles the state of the probe with the state of the
system. Measuring the probe performs an indirect measurement of the
system. From the Stinespring dilation~\cite{Stinespring}, this effectively
implements a completely positive map with Kraus operators given by
$b_j=\bra{j}U\ket{0}$, where $\{\ket{j}\}$ is the computational basis of the
probe.  Outcome $j$ corresponds to the measurement operator $B_j=b_j^\dagger
b_j$ and the conditional post-measurement state is
% $\rho_j=b_j \rho b_j^\dagger /\tr[b_j \rho b_j^\dagger]$.
 \begin{equation}
 \rho_j=\frac{b_j \rho b_j^\dagger}{\tr[b_j \rho b_j^\dagger]}.
 \end{equation}
By choosing suitable unitaries, any
binary outcome POVM can be implemented at each stage.

Conditioned on the result of the first measurement, a second measurement is
performed, a third, and so on (Fig.~\ref{fig:seq}). This builds up a binary
measurement tree with each pair of branches representing a different binary
POVM, depending on the previous results. Each node represents the effective
measurement operator (given by Eq.~\ref{eq:seq}) obtained at that
point. Hence, after $t$ measurements, the effective POVM may have as
many as $N=d^t$ elements at the lowest level for a $d$-level probe.

\section{Binary Measurement Trees}

It is easy to build up POVMs with many elements from a
binary measurement tree. However, given an arbitrary POVM with elements
$M_j$, constructing such a measurement tree which implements it is not so
obvious. Here we show how it may be done.

\begin{figure}
\includegraphics[width=0.45\textwidth]{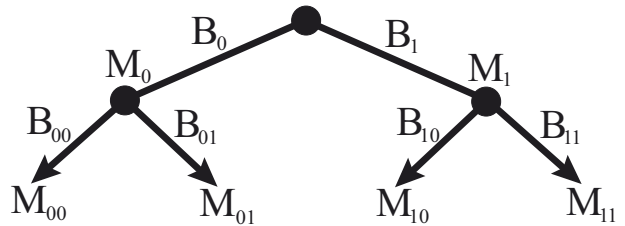}
\caption{Four-outcome POVM. The effective measurement operators $\{M_{ij}\}$
  are given by the $B$ measurements at each step. We want to determine what
  binary measurements $B$ are required to implement $\{M_{ij}\}$.}
\label{fig:fourPOVM}
\end{figure}

It is instructive to look at the simplest non-trivial binary POVM tree with
$t=2$ (Fig.~\ref{fig:fourPOVM}). Let $B_i$ and $B_{ij}$ denote the binary
measurement operators performed at the first and second stage, and $M_i,
M_{ij}$ denote the cumulative measurement operators.  The following should
hold, where $j,k=0,1$:
\begin{subequations}
\begin{eqnarray}
M_{j}&=&B_{j},\\
\one&=&B_{0}+B_{1}=B_{j0}+B_{j1},\\
M_{j}&=&M_{j0}+M_{j1},\\
m_{jk}&=&b_{jk} m_j.
\end{eqnarray}
\end{subequations}
Let us take $b_0=m_0$, $b_1=m_1$ and use the ansatz
$b_{ij}=m_{ij}\tilde{b}_i^{-1}$ where the Moore-Penrose pseudo-inverse
$\tilde{A}^{-1}$ of an operator $A$ is uniquely defined by~\cite{hornjohnson}
\begin{subequations}
\begin{eqnarray}
A\tilde{A}^{-1}A&=&A,\\
\tilde{A}^{-1}A\tilde{A}^{-1}&=&\tilde{A}^{-1},\\
A\tilde{A}^{-1}&=&(A\tilde{A}^{-1})^\dagger,\\
\tilde{A}^{-1}A&=&(\tilde{A}^{-1}A)^\dagger.
\end{eqnarray}
\end{subequations}
We shall prove that the $b_{ij}$ so constructed correspond to POVM operators
and solve the task.

First, since $M_i$ is a positive operator, $M_i=\sum_{k=1}^r
\lambda_{ik}\proj{e_{ik}}$, with positive eigenvalues
$\lambda_{ik}$ and corresponding eigenvectors $\ket{e_{ik}}$;
$r$ is the rank of $M_i$. The $M_{ij}$ are positive operators
and $\sum_{j} M_{ij}=M_{i}$, so the null space of $M_{i}$ is contained in the
null spaces of $M_{ij}$, hence $M_{ij}=\sum_{k,l=1}^r
\phi_{ij,kl}\op{e_{ik}}{e_{il}}$ for some $\phi_{ij,kl}$. Similarly, $m_{ij}=\sum_{kl=1}^r
\varphi_{ij,kl}\op{e_{ik}}{e_{il}}$ for some $\varphi_{ij,kl}$.

We can expand $b_i=m_i=\sum_k \sqrt{\lambda_{ik}} V_i\proj{e_{ik}}$ for some
unitary $V_i$, similarly $\tilde{b}_i^{-1}=\sum_k
1/\sqrt{\lambda_{ik}}\proj{e_{ik}}V_i^\dagger$. Hence, we can see that
\begin{eqnarray}
(m_{ij}\tilde{b}_i^{-1})b_{i}&=& 
\sum_{kl=1}^r\varphi_{ij,kl}\op{e_{ik}}{e_{il}}
\sum_{s=1}^r\frac{1}{\sqrt{\lambda_{is}}}\proj{e_{is}}V_i^\dagger\times\nonumber\\
&&\sum_{t=1}^r \sqrt{\lambda_{it}} V_i\proj{e_{it}} 
=m_{ij}.\nonumber
\end{eqnarray}
In general, completeness of $\{B_{ij}\}_j$ requires us to modify our
original ansatz by adding an extra operator,
\begin{equation}
b_{ij}=m_{ij}\tilde{b}_{i}^{-1} +a_j g_{i},
\label{eq:soln}
\end{equation}
where $g_{i}=\sum_{j=r+1}^d \proj{e_{ij}}V_i^\dagger$ is an isometry on the
null space of $b_i^\dagger$ and the coefficients satisfy $\sum_j |a_j|^2=1$. We
have defined $g_i$ so that $g_i b_i=0$. With this slight modification, it is
easy to show that $\sum_k B_{jk}=\one$.

\begin{figure}
\includegraphics[width=0.45\textwidth]{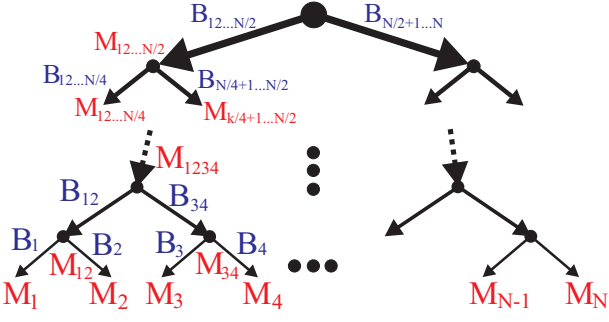}
\caption{Binary POVM Tree. An arbitrary POVM is given by operators
  $M_j=m_j^\dagger m_j$. A sequence of binary outcomes with measurement operators
  $B_\mu,B_\nu,\ldots,B_\xi$ leads to the measurement $M_j=m_j^\dagger
  m_j$ where $m_j=b_\xi \ldots b_\nu b_\mu$.  
  Half of the possible results in the branches
  below are eliminated at each step.}
\label{fig:tree}
\end{figure}

For a general POVM $\{M_k\}$ with $K$ elements, we first pad
the set with null operators until it contains $N$ elements for
$N=2^t,t=\lceil\log_2 K\rceil$ (Fig.~\ref{fig:tree}).
In a convenient change of notation, the cumulative POVM at the $j^\text{th}$
level consists of $2^j$ operators $M_{x}$ where $x$ is a sequence of $2^{t-j}$ 
numbers indicating which of the possible outcomes $M_x=\sum_{i=1}^{2^{t-j}}
M_{x_i}$ sit in the corresponding branches below. A binary POVM
$\{B_{x_a},B_{x_b}\}$ splits each node into two possible branches, each
containing half of the remaining outcomes. We now determine the
binary POVMs $B$ which take us from a higher to lower branch.

At the first level, $B_{12\ldots N/2}=\sum_{i=1}^{N/2}M_i=M_{12\ldots N/2}$
and $B_{N/2\ldots N}=\sum_{i=n/2+1}^N M_i=M_{N/2+1\ldots N}$. At the second
level, from the previous section we have
%\begin{subequations}
\begin{eqnarray}
b_{12\ldots N/4}&=&
m_{12\ldots N/4}\tilde{b}_{12\ldots N/2}^{-1} +g_{12\ldots N/2}\nonumber\\
b_{N/4+1\ldots N/2}&=&
m_{N/4+1\ldots N/2}\tilde{b}_{12\ldots N/2}^{-1} +g_{12\ldots N/2}\nonumber\\
b_{N/2+1\ldots 3N/4}&=&
m_{N/2+1\ldots 3N/4}\tilde{b}_{N/2+1\ldots N}^{-1} +g_{N/2+1\ldots N}\nonumber\\
b_{3N/4+1\ldots N}&=&
m_{3N/4+1\ldots N}\tilde{b}_{N/2+1\ldots N}^{-1} +g_{N/2+1\ldots N}\nonumber
\end{eqnarray}
%\end{subequations}
where we have absorbed the normalization of the $g_x$ operators.
At subsequent levels, we can express the required binary POVMs as
\begin{subequations}
\begin{eqnarray}
b_{x_a}&=&m_{x_a}\tilde{m}_{x_ax_b}^{-1}+g_{x_ax_b}\\
b_{x_b}&=&m_{x_b}\tilde{m}_{x_ax_b}^{-1} +g_{x_ax_b}
\end{eqnarray}
\end{subequations}
where $x_ax_b$ is the concatenation of the strings $x_a$ and $x_b$. At the last 
level $b_1=m_1 \tilde{m}_{12}^{-1}+g_{12}$ and $b_2=m_2 \tilde{m}_{12}^{-1}+g_{12}$. 
Note that the unitary freedom $m_x\rightarrow V_j m_j$ leaves the observed probabilities
invariant but simply rotates the post-selected states after each measurement.

For an $N$ element POVM, we need only a probe qubit and $\lceil
\log_2{N}\rceil$ rounds of binary measurements. Let us determine the number of
operations required to implement this measurement compared to other
methods. For a measurement with $N$ outcomes on a $d$-dimensional quantum
system, the standard Neumark extension requires a $N\times N$ unitary
transform.  This can be realized with $N(N-1)/2$ operations between pairs of
basis states~\cite{reck}, followed by a projective measurement in
the $N$-dimensional space. The realization using just a single extra degree of
freedom \cite{WY2006} requires a $(d+1)\times (d+1)$ unitary transform to be
implemented a maximum of $N-d$ times, giving in total a maximum of
$(N-d)(d+1)d/2$ operations~\footnote{If all outcomes are equally likely the
  average number operations is half the maximum; \textit{a priori} knowledge
  of probabilities may reduce the average number of operations.}. The
binary search requires a $2d\times 2d$ transform to be implemented $\lceil
\log_2{N}\rceil $ times, that is, $\lceil \log_2{N}\rceil d(2d-1)$ pairwise
interactions, a significant speedup if $N$ is large.

\section{Example: Tetrad Measurement}

\begin{figure}
\includegraphics[width=0.45\textwidth]{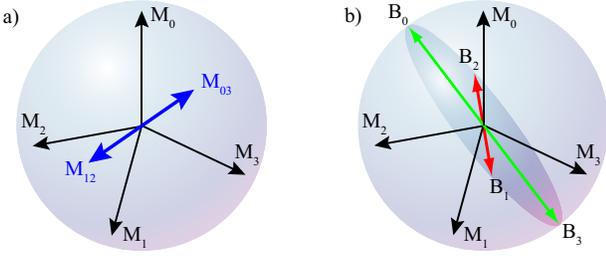}
\caption{Tetrad POVM. a) The first binary measurement is a
  partial filtering with operators $\{M_{03},M_{12}\}$. b) The second
  (projective) measurements depend on the outcome of the first measurement. 
  If $M_{03}$ ($M_{12}$) was obtained, then $\{B_0,B_3\}$
  ($\{B_0,B_3\}$) is measured. The second binary measurements are projective measurements
  in the plane perpedicular to the directions of the first measurement and the
  $B_j$ lie in the direction of the projection of the $M_j$ upon this plane.}
\label{fig:tetrad}
\end{figure}

As an example of the method, consider the symmetric informationally complete
POVM of a single qubit, the so-called tetrad measurement, with measurement
operators $M_j=\proj{\psi_j}$ given by~\cite{singapore}
\begin{subequations}
\begin{eqnarray}
\ket{\psi_0}&=&\frac{1}{\sqrt{2}}\ket{0},\\
\ket{\psi_1}&=&\frac{1}{\sqrt{6}}(\ket{0}+\sqrt{2}e^{\frac{2\pi i}{3}}\ket{1}),\\
\ket{\psi_2}&=&\frac{1}{\sqrt{6}}(\ket{0}+\sqrt{2}e^{\frac{4\pi i}{3}}\ket{1}),\\
\ket{\psi_3}&=&\frac{1}{\sqrt{6}}(\ket{0}+\sqrt{2}\ket{1}).
\end{eqnarray}
\end{subequations}
Although the tetrad POVM can be performed in one
projective step with the addition of just one extra qubit, it is instructive
to demonstrate the binary tree approach using this example.

At the first stage, we are free to choose which two final measurement
operators to group together, for instance,
\begin{subequations}
\begin{eqnarray}
M_{03}=B_{03}=M_0+M_3
=\frac{1}{3}\left(
\begin{array}{cc}
1 & \frac{1}{\sqrt{2}} \\
\frac{1}{\sqrt{2}} & 2
\end{array}
\right),\\
M_{12}=B_{12}=M_1+M_2
=\frac{1}{3}\left(
\begin{array}{cc}
2 & \frac{-1}{\sqrt{2}} \\
\frac{-1}{\sqrt{2}} & 1
\end{array}
\right).
\end{eqnarray}
\end{subequations}
We are also free to choose the Kraus operators $m_{x}=\sqrt{M_{x}}$
using the singular value decomposition. For example,
\begin{subequations}
\begin{eqnarray}
m_{03}&=&\sqrt{\lambda_{+}}\proj{e_{+}}+\sqrt{\lambda_{-}}\proj{e_{-}},\\
m_{12}&=&\sqrt{\lambda_{-}}\proj{e_{+}}+\sqrt{\lambda_{+}}\proj{e_{-}},
\end{eqnarray}
\end{subequations}
where the eigenvalues and eigenvectors are
\begin{subequations}
\begin{eqnarray}
\lambda_{\pm}&=&(1\pm\/\sqrt{3})/2,\\
\ket{e_{\pm}}&=&(\pm\sqrt{3\pm\sqrt{3}}\ket{0}+\sqrt{3\mp\sqrt{3}}\ket{1})/\sqrt{6}.
\end{eqnarray}
\end{subequations}

Although $M_{03}$ and $M_{12}$ share their eigenbases, we need a full $4\times
4$ Neumark extension binary POVM so that the post-measurement state is ready
for the next stage. We couple the system via $U$ to a auxiliary probe qubit
prepared in the state $\ket{0}_{a}$. Then, projecting the probe onto states
$\ket{0}_{a}$ and $\ket{1}_{a}$ corresponds to operations $_{a}\langle 0|
U |0\rangle_{a} = m_{03}$ and $_{a}\langle 1| U |0\rangle_{a} = m_{12}$
on the system. A suitable coupling $U$ is constructed by making a $4\times 4$
Neumark extension of the two-column matrix with its first two rows given by
$m_{03}$, and last two rows by $m_{12}$.  In the basis
$\{\ket{e_{\pm}}\ket{j}_{a}\}$, one possible $U$ is
\begin{equation}
U=\left(\begin{array}{c c c c}
    \sqrt{\lambda_+} & 0 &\sqrt{\lambda_-} & 0\\
    0                            & \sqrt{\lambda_-} & 0     & \sqrt{\lambda_+}\\
    \sqrt{\lambda_-} & 0 & -\sqrt{\lambda_+} & 0\\
    0 & \sqrt{\lambda_+} &0 & -\sqrt{\lambda_-} \end{array}\right).
\end{equation}

In this example, the positive operators $m_{jk}$ are invertible so the $b_j$
for the next step are easily obtained as
\begin{subequations}
\begin{eqnarray}
b_0 &=& \sqrt{M_0}\sqrt{M_{03}}^{-1},\quad b_1=\sqrt{M_1}\sqrt{M_{12}}^{-1}\\
b_2 &=& \sqrt{M_2}\sqrt{M_{12}}^{-1},\quad b_3=\sqrt{M_3}\sqrt{M_{03}}^{-1},
\end{eqnarray}
\end{subequations}
which gives
\begin{eqnarray}
B_0&=&b_0^\dagger b_0=\frac{1}{2}\left(
\begin{array}{cc}
1+\sqrt{\frac{2}{3}} & \frac{-1}{\sqrt{3}}\\
\frac{-1}{\sqrt{3}} & 1+\sqrt{\frac{2}{3}}
\end{array}
\right),\\
B_1&=&b_1^\dagger b_1=\frac{1}{2}\left(
\begin{array}{cc}
1 & -i\\
i & 1
\end{array}
\right),\\
B_2&=&b_2^\dagger b_2=\frac{1}{2}\left(
\begin{array}{cc}
1 & i\\
-i & 1
\end{array}
\right),\\
B_3&=&b_3^\dagger b_3=\frac{1}{2}\left(
\begin{array}{cc}
1-\sqrt{\frac{2}{3}} & \frac{1}{\sqrt{3}}\\
\frac{1}{\sqrt{3}} & 1-\sqrt{\frac{2}{3}}
\end{array}
\right).
\end{eqnarray}
The $b_j$ are rank one operators but are not Hermitian. We can visualize the
sequence of measurements on the Bloch ball (Fig.~\ref{fig:tetrad}).

\section{Conclusion}

In conclusion, we provide a constructive proof of the universality of sequential
two-outcome POVMs. We show how to construct binary measurement trees to
implement any generalized measurement through a sequence of indirect binary
POVMs requiring only an extra auxiliary qubit. This avoids having to 
manipulate extended Hilbert spaces (larger than twice the dimension of the
measured system) and reduces the number of required operations when the number
of outcomes becomes large. The number of steps is logarithmic in the number of
measurement outcomes. The required interaction to perform binary POVMs exists
in physical systems such as cavity quantum electrodynamics
(CQED)~\cite{guerlin2007} where the state of a field can be probed by an
atom-cavity interaction and the atom measured. So far, projective measurements
have been performed with a fixed interaction and measurement, but it should be
possible with feed-forward and suitable control fields to implement a full
POVM measurement as described here.

%The result also
%has implications for the divisibility~\cite{wc2006} of arbitrary completely
%positive maps with the addition of classical feed-forward and suggests a rich
%hierachy of non-Markovianity.

\begin{acknowledgments}
DKLO acknowledges the support of the Scottish Universities Physics Alliance (SUPA).
EA gratefully acknowledges the support of the Royal Society of London. We
thank S. G. Schirmer for valuable discussion.
\end{acknowledgments}

%%%%%%%%%%%%%%%%%%%%%%%%%%%%%%%%%%%%%%%%%%%%%%%%%%%%%%%%%%%%%%%%
%%\section{References}

\end{document}